\documentstyle[12pt,a4]{article}
\begin{document}
\def\ve{\vfil\eject}
\begin{flushright}
TIT/HEP--250/COSMO--43 \\
March, 1994
\end{flushright}

\title{Quantized Topological 2-form Gravity
}
\author{Tomoyuki Inui, \ Akika Nakamichi}
\date{Department of Physics, Tokyo Institute of Technology \\
Oh-okayama, Meguro-ku, Tokyo 152 JAPAN}
\maketitle
\begin{abstract}
\par

 We develop a perturbation theory of four-dimensional
topological 2-form gravity without cosmological
constant.
 A 2-form and an $SU(2)$ connection
1-form are used as fundamental variables instead of metric.
 There is no quantum correction from
two-loop and higher orders in covariant gauge, in Landau gauge
and in background covariant gauge.
 We improve naive dimensional regularization and
calculate an  exact quantum correction at one-loop order.
 We observe that the renormalizability depends on
gauge choice in topological 2-form gravity.
It is unrenormalizable in the covariant gauge and in the
background covariant gauge.
 On the other hand in the Landau gauge,
we obtain finite but $SU(2)$ non-covariant quantum corrections.
 There may exist anomaly for the $SU(2)$ gauge symmetry.

(Talk given by A. Nakamichi at the Workshop on General
Relativity and Gravitation held at Tokyo University,
January 17-20,1994.)
\end{abstract}

\section{Introduction}
\par
        How to describe the four dimensional quantum gravity is
a longstanding problem in theoretical physics.
 To construct a renormalizable gravity theory in four
dimensions, it seems that
we need new variables or new actions
instead of metric or the Einstein action.
 One of the candidates of
the   new variable, self-dual 2-form, was introduced in
in the early 1960s~\cite{Isra}~\cite{Ple}.
 We will refer to it as 2-form Einstein gravity.
      However the unrenormalizability of the theory still
 remains as a major problem.
  To avoid renormalizability problem,
topological gravity has studied, which is apparently  free from divergence.
  Recently a close connection between
 2-form Einstein gravity and
topological gravity was shown~\cite{Horo}~\cite{Lee}.
 The action of topological 2-form gravity is obtained by dropping a
certain constraint term in the action of 2-form Einstein gravity.
 This constraint term is proportional to the Weyl tensor, which governs
the modes of the gravitational waves.
 Thus topological 2-form gravity is interpreted as
a theory which can describe the
vacuum state (i.e. the state of no gravitational waves)
of the Einstein gravity.
 Since we would like to construct a renormalizable theory of
gravity and demonstrate its finiteness in a simpler model,
we consider topological 2-form gravity in this paper.
  In the near future, we will be able to attack the 2-form Einstein
gravity to make good use of this experience.

\section{Review of 2-form Einstein Gravity}%

         The action of the Euclidean\footnote{In case of
 Lorentzian signature, we need reality condition to
recover real metric.
 Therefore in this paper we will work in Euclidean signature
for simplicity.}
 2-form Einstein gravity
 is given in terms
of a 2-form $\Sigma^k$ and an $SU(2)$ spin
connection 1-form $\omega_k$ in the presence of the cosmological
constant $\lambda$,
\begin{equation}
 S = \int {\Sigma^k} \wedge {R_k}
  + \lambda  \  {\Sigma^k} \wedge {\Sigma_k}
  + {1 \over 2} \  {\psi_{ij}}{\Sigma^i} \wedge {\Sigma^j} \ ,
\label{eq:Einstein}
\end{equation}
where $R_k \equiv d\omega_k + ( \omega \times \omega )_k $ and
$ \psi_{ij} $ is a  symmetric trace-free Lagrange multiplier field.
Here we use the notation for $SU(2)$ indices
$ ( F \times G )_i = \epsilon_{ijk} F^j G^k$ and
   $\epsilon_{ijk}$ is the structure constant of $SU(2)$.
 The $SU(2)$ indices $i, j, k, ... = 1,2,3$ imply that they transform
under {\it chiral} local-Lorentz representation (1, 0)
of $ SO(4) = SU(2) \times SU(2) / {\bf Z}_2 $.
 In this formulation, the metric field $g_{\mu \nu}$ is defined
in terms of the 2-form $\Sigma^k$ as
\begin{eqnarray}
g^{1\over 2} g_{\mu \nu}
  &=& {1 \over 12} \  {\varepsilon^{\alpha \beta \gamma \delta}}
       \   {\Sigma_{\mu \alpha}}^k
    \    ( \Sigma_{\beta \gamma } \times
    \Sigma_{\delta \nu } )_k , \nonumber \\
  g &{\equiv}& det(g_{\mu \nu})\ . \label{eq:metric}
\end{eqnarray}
 The constraint equation, which is obtained
 by varying eq.(\ref{eq:Einstein}) with respect to
$\psi_{ij}$ , implies that $\Sigma^{k}$  is composed
as a wedge product of 1-forms $e^k$ :
\begin{equation}
 {}^{( trace-free )} \Sigma^{(i} \wedge \Sigma^{j)}=0
\iff
 {\Sigma_{\mu \nu}}^k =
- {\varepsilon_{\mu \nu \lambda \rho}}
   \, e^{\lambda a} \, e^{\rho b} \, {\eta^k}_{ab}
\label{eq:constraint}
\end{equation}
here ${\eta^k}_{ab}$ is t'Hooft's $\eta$ symbol, which
connects $SU(2)$ index $k$ and $SO(4)$ indices $a, b$ and
self-dual for $SO(4)$ indices~\cite{t'Hooft}.
 Using this result and translating the $SU(2)$ spinor indices into the
$SO(4)$ (the local-Lorentz) indices,
 the action (\ref{eq:Einstein}) becomes identical with
the chiral decomposition of the first-order Palatini action
 in which the usual spin connection is replaced by
its anti-self-dual
part with respect to the $SO(4)$ indices.
 As is discussed in~\cite{JacoSmo},
this chiral action also gives the Einstein equation.
 With the help of the equations of motion derived from
the action (\ref{eq:Einstein}),
  this $\psi_{ij}$ is determined to be
proportional to the anti-self-dual part of the Weyl tensor which just governs
the modes of the gravitational waves.
\par
 Since the action (\ref{eq:Einstein}) describes general relativity,
it is invariant under the local-Lorentz transformation
and diffeomorphism,
\begin{eqnarray}
  \delta \omega^i &=&  D\phi^i + {\cal L}_\xi \omega^i \ ,
    \nonumber \\
  \delta \Sigma^i &=& \epsilon_{ijk} \Sigma^j \phi^k
     + {\cal L}_\xi \Sigma^i    \ ,
\end{eqnarray}
where ${\cal L}_\xi$ is the Lie derivative with respect to a vector
field $\xi^\mu$, and the chiral local-Lorentz transformation corresponds to
the $SU(2)$ gauge transformation with a parameter $\phi^k$.
 We use the following notation for
 spin-covariant exterior derivative $D$ :
$ D\phi^i \equiv d\phi^i +
2 \epsilon_{ijk} \omega^j  \phi^k  $ ,
in this and the next section.

\section{Classical Topological 2-form Gravity}    %

      Let us consider the situation when we drop the last
constraint term (\ref{eq:constraint}) in the action
(\ref{eq:Einstein}).
 Then the action becomes
\begin{equation}
  S_{classical}= \int \Sigma^k \wedge R_k
                   + \lambda \ \Sigma^k \wedge \Sigma_k \ .
 \label{eq:topaction}
\end{equation}
 In this particular case, a new symmetry with a parameter 1-form
$\theta^k$ emerges in addition to the chiral local Lorentz symmetry,
which we will call as $\theta_1 -$ symmetry,
\begin{eqnarray}
  \delta \omega^i &=& D\phi^i - 2 \lambda \theta^i \, ,  \nonumber \\
  \delta \Sigma^i &=& 2 \epsilon_{ijk} \Sigma^j \phi^k
      + D\theta^i \ .
 \label{eq:topsymm}
\end{eqnarray}
 Although the theory remains invariant under diffeomorphism just as
in the previous section, there is no need to add it
to eq.(\ref{eq:topsymm}) because, modulo the equations of motion derived from
(\ref{eq:topaction}),
diffeomorphism  can be generated by the
combination of the above chiral local-Lorentz
and $\theta_1 -$transformations (\ref{eq:topsymm}) with
$\phi^k
= \xi^\nu \omega_\nu^k$ and ${\theta_\mu}^k
= 2\xi^\nu \Sigma_{\nu \mu}^k$.
 With the appearance of the $\theta_1 -$ symmetry, the theory turns out
to be on-shell reducible in the sense that
the classical gauge transformation laws (\ref{eq:topsymm})
are invariant under
\begin{eqnarray}
    \delta \phi^k &=&  2 \lambda \upsilon^k \ , \nonumber \\
    \delta \theta^k &=& D \upsilon^k \ ,
 \label{eq:reduc}
\end{eqnarray}
if the equations of motion are satisfied.

\section{An Improved Dimensional Regularization}

We have to calculate some superficially divergent
Feynman integrals in order to demonstrate that topological
2-form gravity theory is finite.
To obtain  finite quantum corrections in a perturbation
theory, we need some kind of regularization.
 It seems that the best regularization scheme is
the Pauli-Villars regularization~\cite{Pauli},\cite{Bell},
 in which the gauge invariance is manifest.
 However in our model, owing to the existence of unusual
2-form ${\Sigma_{\mu \nu}}^k$, we do not know how to
introduce regulator fields.
 We use an improved dimensional regularization
 scheme in this paper.
 The point is that the dimensions
of the  symbol $\varepsilon^{\mu \nu \lambda \rho}$
is fixed to be four.
 \par
 We are confronted by difficulty:
In the quantum action,
the field ${\Sigma_{\mu \nu}}^k$ in the classical term
remains four-dimensional, while ${\Sigma_{\mu \nu}}^k$
 in the gauge-fixing term is extended to be $n$-dimensional.
 The same situation occurs in the field ${\omega_\mu}^k$.
 Then we cannot obtain the propagators in the extended
dimensions.
 Thus our improved regularization scheme is not justified
in the extended dimensions.
\par
 Putting aside the above difficulty, we will proceed to
carry out our calculation by our improved
dimensional regularization.

\section{Topological 2-form Gravity in Covariant Gauge}

        In the topological theories, metric is needed
only for the gauge fixing.
 We choose the flat Euclidean metric as a background
metric.
 Owing to the reducibility, we use powerful BFV Hamiltonian
formalism\footnote{We  also
 used BV Lagrangian formalism~\cite{BV} for the check
of our calculation; both results agree with each other.}
  for the topological 2-form gravity, which was carried out
in Ref.\cite{Lee}.

 So far we have obtained quantum tree action and BRST transformations
in the presence of cosmological constant $\lambda$.
 Although we will be able to
 proceed in this general model,
 we would like to proceed in a simpler model.
 Setting  $\lambda = 0$ , now the classical
symmetry becomes
\begin{eqnarray}
  \delta \omega^i &=& D\phi^i \, , \nonumber \\
  \delta \Sigma^i &=& 2 \epsilon_{ijk} \Sigma^j \phi^k
      + D\theta^i \ .
\end{eqnarray}
  The on-shell reducibility becomes
\begin{eqnarray}
    \delta \phi^k &=&  0 \ , \nonumber \\
    \delta \theta^k &=& D \upsilon^k \ .
\end{eqnarray}

 We have chosen the gauge fixing conditions for
chiral local-Lorentz, $\theta_1-$symmetry  and on-shell
reducible symmetries\footnote{
 Should we fix such the on-shell symmetry?
 The answer is `Yes'.
 If we do not fix it, we cannot construct propagators.},
 respectively:
\begin{eqnarray}
  \partial^\mu \omega_{\mu i} &=& 0 \ , \nonumber \\
  D^\nu \Sigma_{\mu \nu i} &=& 0 \ , \nonumber \\
  D^\mu C_{\mu i} &=& D^\mu {\bar C}_{\mu i} = 0 \ ,
\label{eq:covagauge}
\end{eqnarray}
where ${C^\mu}_i$ and ${\bar C}^\mu_i $ are fermionic ghost and
anti-ghost, respectively, for ${\theta_\mu}^k$
parameter in $\theta_1-$symmetry.
 According to BFV quantization method we enlarge the
phase space as follows.
 We introduce fermionic ghost $C^k$ and anti-ghost
$\bar{C}^k$ for $\phi^k$ parameter in chiral local Lorentz
symmetry and bosonic ghost-for-ghost ${C_1}^k$ and its
anti-field ${\bar{C}_1}^k$ for $v^k$ parameter
in the reducible symmetry.
 The subscript $1$ means the class of the first stage
in the gauge structure.
 We also introduce Lagrange multiplier fields:
bosonic $\pi^k$ for the gauge fixing of chiral local
Lorentz, bosonic ${\hat{\pi}_\mu}^k$ for that of
$\theta_1-$symmetry, and
fermionic ${\hat{\pi}_1}^k$ and $\tau^k$ for that of
reducible symmetries.
 Besides these fields, we need bosonic extra-ghost $\rho^k$
in order to make the theory covariant completely.

        Using the BRST invariant action,
 propagators can be written down.
 All the fields in  this theory
cannot freely propagate; they necessarily change
into different fields in the course of propagation.

 There has been an expectation that
 effects of higher order loops
do not contribute in topological field theories in general.
 As a matter of fact in the three-dimensional non-Abelian BF theory
and Chern-Simons theory, it is proven that there exist at most
one$-$loop corrections~\cite{Oda}.
 In this analogy we are naturally led to the following
conjecture:
There exist at most one$-$loop
quantum corrections in Topological 2-form Gravity in
the covariant gauge.

 We have proved that this is the truth and obtained
the following relation:
\begin{equation}
 L - 1 + E_\omega + E_{\hat{\pi}_\mu} +
  E_{\hat{\pi}_1} + E_{\bar{C}_\mu}
  + E_{\bar{C}_1} + E_C = 0 \, .
 \label{eq:starstar}
\end{equation}
 Here we denote
the number of $Y-$ field external lines in a connected Feynman diagram
$H$ as $E_Y$,
 the number of loops
in $H$ as $L$.
 Since the number of each external line is not less than 0,
there is no connected diagram whose number of loops is
 greater than one.
 We can get the exact result
from at most one-loop calculation.

 Now we study the one-particle-irreducible ( 1PI )
connected diagrams with
 external lines amputated.
 Since we are interested in the vertex functions,
the important fields are
 connected with the amputated external lines.
 Now only ${\omega_\mu}^i$ and
${\hat{\pi}_\mu}^i$ are such fields
at one loop order.
 After integrating out dynamical fields
in the effective action, there remain only these two fields
which appear in quantum corrections to the action.
 We denote the number of these fields as
$X_\omega$ and $X_{\hat{\pi}_\mu}$, respectively
in a 1PI diagram, say $H$.
 The superficially divergent graphs are
\begin{equation}
( X_\omega ,  X_{\hat{\pi}_\mu} ) =
 4 - ( X_\omega +  X_{\hat{\pi}_\mu} )
\end{equation}

 We obtain one-loop quantum corrections
of superficially divergent graphs.
 We use `background field method' for this calculation.
 In the `background field method', we decompose
amputated external fields
into background c-number fields plus quantum q-number
fields:
\begin{eqnarray}
 {\omega_\mu}^k  && \rightarrow \;  {\omega_\mu}^{k (q)}
      + \widetilde{{\omega_\mu}^k } \, ,
 \nonumber \\
 {\hat{\pi}_\mu}^k && \rightarrow \;
     {\hat{\pi}_\mu}^{k (q)}
     + \widetilde{{\hat{\pi}_\mu}^k} \, .
\end{eqnarray}
 Here the superscript $(q)$ in the right hand sides means
 quantum fields.
The tilde over the fields in the right hand sides
means that  they are the background fields.
 We will omit the quantum label $(q)$ from now on.

 In the background field method, we can calculate
the one-loop quantum corrections
using only the terms which are bilinear in
the quantum components.
 Combining these terms,
we construct the superficially divergent diagrams.

First we consider the graph
$( X_\omega \, , X_{\hat{\pi}_\mu} ) = ( 0 , 2 ) $ .

 Its quantum correction is,
\begin{equation}
 - \Gamma_{\hat{\pi}_\mu \hat{\pi}_\mu}
  = - 2 \widetilde{{\hat{\pi}_\mu}^i}
     \widetilde{{\hat{\pi}_\rho}^i}
  \int \frac{d^4 p}{(2 \pi)^4}
  \frac{\delta_{\mu \rho} p_\alpha (p + k)^\alpha
         - p^\rho (p + k)^\mu}{p^2 (p + k)^2}
   \, . \label{eq:divergence}
\end{equation}
 This quantum correction yields quadratic divergence.
 In order to cancel this divergence, we can think of
adding the bilinear term  of $\hat{\pi}_\mu$,
which is constructed from adding this field to the gauge fixing
condition for ${\theta_1}-$symmetry.
 However this term cannot be arranged to cancel
eq.(\ref{eq:divergence}).
Thus we conclude that we cannot renormalize this divergence.
   We consider that this unrenormalizable divergence is
because of our gauge choice,
 and investigate this problem further in the next section.

\section{Topological 2-form Gravity in Landau Gauge}

        Why there appeared unremovable divergence in four-dimensional
topological 2-form gravity in the previous chapter?
 So far it has been believed that all the topological
field theories are finite.
 In fact in
 the case of Schwartz type topological theories one can
prove the finiteness of the theory from the fact that
the space of
solutions to the field equations, modulo the gauge symmetries,
is finite dimensional.
However the proof of finiteness applies
 only when the theory is
restricted to an appropriate moduli space that the finiteness
of the model is manifest.
 As we will see in this section,
 this restriction to a finite dimensional moduli space
is nothing but a gauge choice.

  Therefore we must choose alternative gauge fixing conditions
 to avoid the divergence in the two$-$point function of
the Lagrange multiplier field $\hat{\pi}_\mu$.
 The essential observation is that in the
 Landau gauge, this field cannot enter in the
three-point vertex.
 This fact strongly suggests us to use the Landau gauge.
 There is another reason to choose the Landau gauge.
 In the course of the calculations in the covariant gauge,
we noticed the papers~\cite{Sore1} \cite{Sore2}
 in which the finiteness of four dimensional
 BF theory in the Landau gauge was shown by using Slavnov identity.

        From now on we use the following definition of
covariant derivative and curvature for simplicity,
\begin{eqnarray}
 (D_\mu \theta )^i &=& \partial_\mu \theta^i + \epsilon^{ijk}
    {\omega_\mu}^j \theta^k  \nonumber \\
 {R_{\mu \nu}}^i &=& \partial_\mu {\omega_\nu}^i
                     - \partial_\nu {\omega_\mu}^i
       + \epsilon^{ijk} {\omega_\mu}^j {\omega_\nu}^k
\end{eqnarray}

 Now we choose the gauge fixing conditions for chiral
local-Lorentz, $\theta_1-$symmetry and reducible symmetries,
respectively, in the following Landau gauge:
\begin{eqnarray}
  \partial^\mu \omega_{\mu i} &=& 0 \ , \nonumber \\
  \partial^\nu \Sigma_{\mu \nu i} &=& 0 \ , \nonumber \\
  \partial^\mu C_{\mu i} &=& \partial^\mu \bar{C}_{\mu i} = 0 \ .
\label{eq:langauge}
\end{eqnarray}
  The quantization of this model,
owing to its reducibility,
is not straightforward and requires the BFV or BV procedure
 in the same way as in the previous section.
 This quantization was carried out in ~\cite{Sore3}.
 Since we would like to use the same propagators in the
previous section, we re-define some fields
so that the kinetic terms are the same as
in "covariant gauge".
 This action is also off-shell invariant under
simpler BRST transformations.
\par
 The loop$-$structures in the theories are expected to
be independent of the gauge$-$choices.
 Therefore in this gauge, we are also led to the
 conjecture:
 There is no quantum correction from two-loop
and higher orders
in Topological 2-form Gravity in the Landau gauge.
 In this gauge, we have also proved that this is the case.
The following equation is obtained:
\begin{equation}
 L - 1 + E_\omega + E_{\hat{\pi}_\mu} +
  E_{\hat{\pi}_1} + E_{\bar{C}_\mu}
  + E_{\bar{C}_1} + E_C = 0 \, .
 \label{eq:exact2}
\end{equation}
The superficial degree of divergence
for the diagram $H$ in the Landau gauge is,
\begin{eqnarray}
 \omega ( H ) &=& 4 - ( X_\omega  + X_{\bar{C}_\mu} )
           - 2 X_{\bar{C}_1} \nonumber \\
 &=& 4 - ( X_\omega + X_C ) \, .
\label{eq:degree3}
\end{eqnarray}

 The exact quantum corrections in the Landau
gauge are the following:
\begin{eqnarray}
 - \ \Gamma &=&  - \frac{1}{96 \pi^2} \int d^4 x
   [ ( \partial^\alpha
    \widetilde{\omega^{\beta i}} )
    ( \partial_\alpha \widetilde{\omega_{\beta i}} )
  + 2  ( \partial^\alpha \widetilde{{\omega_\alpha}^i} )
      ( \partial^\beta \widetilde{\omega_{\beta i}} )
\nonumber \\
& & \ +  \epsilon^{ijk}
      ( \partial^\alpha \widetilde{{\omega^\beta}_i} )
       \widetilde{\omega_{\alpha j}}
       \widetilde{\omega_{\beta k}}
\nonumber \\
 & & - \frac{1}{4}  \widetilde{\omega^{\alpha i}}
    \widetilde{\omega_{\alpha i}}
    \widetilde{\omega^{\beta j}}
     \widetilde{\omega_{\beta j}}
 - \frac{1}{2} \widetilde{\omega^{\alpha i}}
     \widetilde{{\omega^\beta}_i}
     \widetilde{{\omega_\alpha}^j}
     \widetilde{\omega_{\beta j}} ]
 \, . \label{eq:kekka}
\end{eqnarray}
  Recall that the field ${\omega_\mu}^i$
 is invariant under the
$\theta_1-$transformation,
when the cosmological constant is 0.
 Therefore our result eq.(\ref{eq:kekka}) is
invariant under the classical  $\theta_1-$symmetry.
 However, this result is not covariant under the
classical $SU(2)$ gauge transformation.
 Thus we observe that the classical $SU(2)$ symmetry
breaks at the quantum level.

\section{Topological 2-form Gravity in
         Background Covariant Gauge}

 In this section, we proceed in the background covariant gauge.
 We use an $SU(2)$ connection as the background field.
 It seems that this gauge is the most appropriate choice for
the perturbation theory in the background field method.
  We choose the gauge fixing conditions for chiral local
Lorentz symmetry, $\theta_1-$symmetry and reducible symmetries,
respectively:
\begin{eqnarray}
 ( \widetilde{D^\mu} {\omega_\mu}^{(q)} )^i &=& 0 \ , \nonumber \\
 ( \widetilde{D^\nu} \Sigma_{\mu \nu } )^i &=& 0 \ , \nonumber \\
 ( \widetilde{D^\mu} C_{\mu } )^i &=&
         ( \widetilde{D^\mu} \bar{C}_{\mu} )^i = 0 \ .
\label{eq:Appegauge}
\end{eqnarray}
 Here the tilde over the covariant derivative  means that
the connection of this covariant derivative $\widetilde{D_{\mu}}$
is the background $SU(2)$ connection
 $\widetilde{{\omega_{\mu}}^i}$.

 We have also proved that there is no quantum
correction from two-loop and higher orders in this gauge.
 We obtain the following equation:
\begin{eqnarray}
 & & L - 1 + E_{\widetilde{\omega}} + E_{\hat{\pi}_\mu} +
  E_{\hat{\pi}_1} + E_{\bar{C}_\mu}
  + E_{\bar{C}_1} + E_C
\nonumber \\
& & \, + \, \frac{1}{2} (
       V_{\Sigma \widetilde{\omega} \widetilde{\omega}}
 + V_{\widetilde{\omega} \widetilde{\omega} \bar{C} C}
 + V_{\widetilde{\omega} \widetilde{\omega} \bar{C}_{\mu} C_{\nu}}
 + V_{\widetilde{\omega} \widetilde{\omega} \bar{C}_1 C_1}
 + V_{\widetilde{\omega} \widetilde{\omega}
             \bar{C}_{\mu} \bar{C}_{\nu} C_1} )
 = 0 \, .
 \label{eq:backL}
\end{eqnarray}
Here we denote the total number of vertices
$\Sigma \widetilde{\omega} \widetilde{\omega}$ in a connected
Feynman diagram as
$V_{\Sigma \widetilde{\omega} \widetilde{\omega}}$.

At one loop order, the superficial degree of divergence for
the diagram $H$ in this gauge is,
\begin{eqnarray}
\omega ( H ) &=& 4 - ( X_{\hat{\pi_\mu}} + 2 X_{\bar{C}_1}
     + X_{\bar{C}_{\mu}} + 2 X_{\hat{\pi}_1}
     + X_{\widetilde{\omega}} )
\nonumber \\
     &=& 4 - ( X_{\hat{\pi_\mu}} + X_C + X_{\hat{\pi_1}}
               + X_{\widetilde{\omega}} )
 \, .
\end{eqnarray}
 The quantum corrections in terms of the background
$SU(2)$ connection $\widetilde{{\omega_{\mu}}^i}$ are:
\begin{eqnarray}
 - \ \Gamma &=& - \ \lim_{n \to 4} \frac{1}{ ( n - 4 ) \, 192 {\pi}^2}
               \int d^4 x \
   [ \  56 \, ( \partial^\alpha
    \widetilde{\omega^{\beta i}} )
    ( \partial_\alpha \widetilde{\omega_{\beta i}} )
\nonumber \\
& & \, \,  - 8 \,  ( \partial^\alpha \widetilde{{\omega_\alpha}^i} )
      ( \partial^\beta \widetilde{\omega_{\beta i}} )
 \ +  132 \,  \epsilon^{ijk}
      ( \partial^\alpha \widetilde{{\omega^\beta}_i} )
       \widetilde{\omega_{\alpha j}}
       \widetilde{\omega_{\beta k}}
 + 73  \, \widetilde{\omega^{\alpha i}}
    \widetilde{\omega_{\alpha i}}
    \widetilde{\omega^{\beta j}}
     \widetilde{\omega_{\beta j}}
\nonumber \\
& & \, \,
  - 46 \, \widetilde{\omega^{\alpha i}}
     \widetilde{{\omega^\beta}_i}
     \widetilde{{\omega_\alpha}^j}
     \widetilde{\omega_{\beta j}} \ ]
\    + \,  {\rm finite \  parts}
 \, . \label{eq:Appekekka}
\end{eqnarray}
 Unfortunately,
this divergence cannot be removed.
 Thus we conclude that the background covariant gauge
is not appropriate in our topological 2-form gravity.

\section{Discussions}

 We have learned that the renormalizability inevitably
 depends on gauge choice in our
topological 2-form gravity.
 Such a gauge-dependence often appears in topological
field theories~\cite{rep}.
 Using  different gauge, but with the
same regulator, yields  different results.
 Therefore  in our topological 2-form gravity,
we also have to choose an appropriate gauge-fixing
conditions in order to restrict the theory
to an appropriate moduli space so that the theory becomes finite.
\par
   The Vilkovisky-De Witt effective
action program gives a notion of a
unique effective action~\cite{Veffect}, ~\cite{rep}.
 It is interesting to apply this program to our model.
Then we may be able to show that
the unique effective action
in our model is the one for the Landau gauge.
 Moreover
we would like to investigate the geometrical
meaning of the Landau gauge.
\par

 Our improved dimensional regularization is
invariant under the $SU(2)$ transformation.
 However our regularization is justified
only in four dimensions.
 Therefore one of the possible origins of
$SU(2)$ non-covariance in the quantum corrections
is our dimensional regularization.

 The details of this talk will be reported in Ref.~\cite{Inui}.

\end{document}